**Experimental Realization of a Three-Dimensional Dirac Semimetal**

Sergey Borisenko[1], Quinn Gibson[2], Danil Evtushinsky[1], Volodymyr Zabolotnyy[1*], Bernd Büchner[1,3], Robert J. Cava[2]

[1] *Institute for Solid State Research, IFW Dresden, P. O. Box 270116, D-01171 Dresden, Germany*

[2] *Department of Chemistry, Princeton University, Princeton, New Jersey 08544, USA*

[3] *Institut fur Festkörperphysik, Technische Universität Dresden, D-01171 Dresden, Germany*

**The three dimensional (3D) Dirac semimetal, which has been predicted theoretically [1], is a new electronic state of matter. It can be viewed as 3D generalization of graphene, with a unique electronic structure in which conduction and valence band energies touch each other only at isolated points in momentum space (i.e. the 3D Dirac points), and thus it cannot be classified either as a metal or a semiconductor. In contrast to graphene [2], the Dirac points of such a semimetal are not gapped by the spin-orbit interaction and the crossing of the linear dispersions is protected by crystal symmetry [1]. In combination with broken time-reversal or inversion symmetries, 3D Dirac points may result in a variety of topologically non-trivial phases with unique physical properties [1, 3-5]. They have, however, escaped detection in real solids so far. Here we report the direct observation of such an exotic electronic structure in cadmium arsenide ($Cd_3As_2$) by means of angle-resolved photoemission spectroscopy (ARPES). We identify two momentum regions where electronic states that strongly disperse in all directions form narrow cone-like structures, and thus prove the existence of the long sought 3D Dirac points. This electronic structure naturally explains why $Cd_3As_2$ has one of the highest known bulk electron mobilities [6]. This realization of a 3D Dirac semimetal in $Cd_3As_2$ not only opens a direct path to a wide spectrum of applications, but also offers a robust platform for engineering topologically-nontrivial phases including Weyl semimetals and Quantum Spin Hall systems.**

Without considering any aspects of topology, electronic band theory classifies materials as insulators, semiconductors and metals depending on the presence and the size of the energy gap between the valence and conduction bands. Semimetals belong to a "buffer zone" between semiconductors and metals and are characterized by a small energy overlap between the conduction and valence bands. This overlap is vaguely defined, partly because the notion of momentum is traditionally not included in the simple classification. In contrast, 3D Dirac semimetals are precisely defined in terms of the momentum dependent band structure: the mentioned overlap occurs only at a set of isolated points in momentum space where linear energy vs. wavevector dispersions cross at the Fermi level. Such an anomalous Fermi surface in three dimensions would obviously be a source of intriguing physical properties, as is the case with its 2D analog – graphene. Considering a 3D Dirac point as an overlapping of two Weyl points [1, 3-5] immediately promotes 3D Dirac semimetals to one of the most wanted new materials from a theoretical point of view: apart from being natively topologically non-trivial, in some cases they can also be viewed as parent materials for accessing the Weyl semimetal state [7].

In order to identify a 3D Dirac semimetal experimentally, one needs to locate a Dirac point in a 3D momentum space and measure the corresponding dispersions of the electronic states. ARPES is an ideal tool for such a search (see Methods Summary), and we apply the sub-Kelvin ultra-high resolution version of this method in the present work [8].

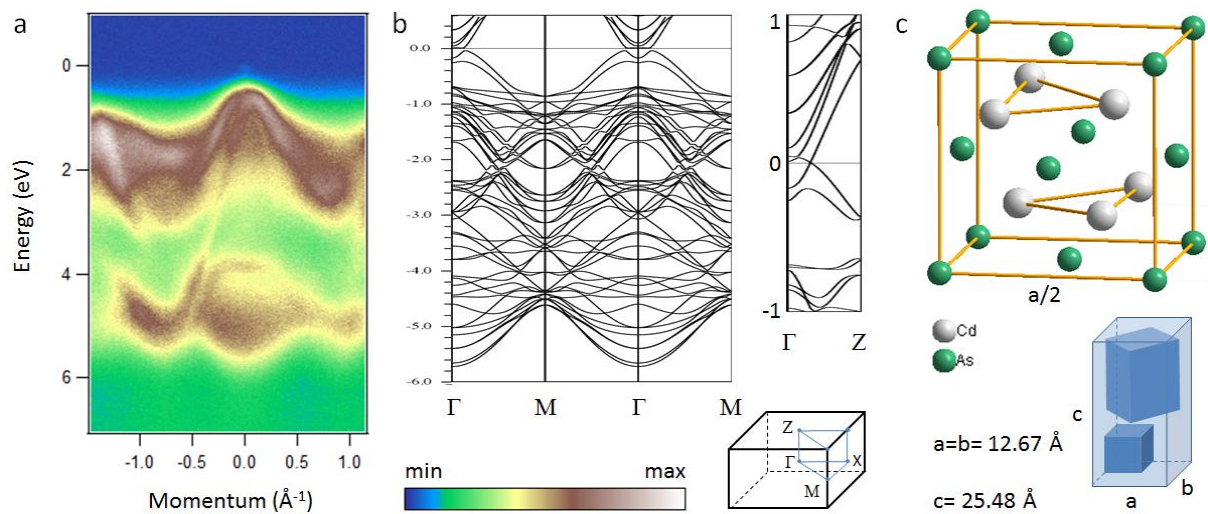

**Figure 1. Electronic and crystal structure of Cd$_3$As$_2$.** (a) Photoemission intensity recorded approximately along the ΓM direction using 90 eV photons. The number of visible features is smaller than suggested by the calculations (b), which is explained by the smaller parent unit cell (c) and matrix elements effects (see SI for more details). Tetragonal BZ and unit cells are shown in the insets. The parent cubic unit cell has random Cd vacancies.

Cd$_3$As$_2$ crystallizes in a distorted relative of the anti fluorite structure, with Cd in tetrahedral coordination and ordered Cd vacancies [9]. It goes through several phase transitions on cooling, which involve Cd vacancy ordering and small structural distortions, and at 225 °C it undergoes a transition from a primitive centrosymmetric tetragonal cell (space group P4$_2$/nmc) to a body centered tetragonal non-centrosymmetric unit cell (space group I4$_1$cd) [10]. No further structural transitions have been observed at lower temperatures. The smaller, centrosymmetric unit cell was used for the electronic structure calculations (see inset to Fig.1 and the supplementary information (SI)), which still capture the main features of the electronic structure of cadmium arsenide, including the existence of a single pair of 3D Dirac points along the ΓZ-direction in the Brillouin Zone. The theoretical study of Wang et. al. [11] also shows these 3D Dirac points in both the centrosymmetric and non centrosymmetric structures, and also predicts Cd$_3$As$_2$ to be topologically nontrivial. In Fig. 1 we compare the results of these relativistic band structure calculations (see Methods Summary) with the data from the ARPES experiment. The typical experimental dataset shown represents the valence band of Cd$_3$As$_2$, the bandwidth of which is of the order of 4.5 eV and is in good correspondence with the calculations. Taking into account that the experimental data in Fig. 1a are taken not exactly along the high-symmetry MΓM-line, the general agreement between the observed dispersion features and the calculated ones is very good. The data show that the ARPES signal is effectively defined by a still smaller crystallographic parent subcell of Cd$_3$As$_2$ – a usual situation when the modulating potential of a lattice distortion leading to a supercell is not very strong and the parent structure is still dominant [12]. This is exactly the case here, as the true unit cell can be viewed as a 2x2x4 supercell of the parent structure subcell shown in Fig. 1c, or as a √2x√2x2 supercell of the tetragonal unit cell used in the calculations. All three structural units are shown schematically in the inset (see also Fig. S1).

In the region between 0.5 eV and the Fermi level, in agreement with calculations, the photoemission intensity is very weak; only close to the Γ-point is there a significant amount of intensity, as is expected in the case of strong dispersion along $k_z$; this is because ARPES is not a bulk sensitive technique and its resolution along $k_z$ is defined by the escape depth. Upon closer inspection of the data, one can notice a weak signal connecting the apparent top of the valence band and the Fermi level. Can this be a signature of the sought connection to the conduction band? As is seen in Fig.1 c, the band structure calculations predict 3D Dirac points located along the ΓZ direction. This means that in order to find the 3D Dirac point, it is not sufficient to explore the $k_x$ - $k_y$ plane at a given $k_z$, and a scan of the excitation photon energy is required.

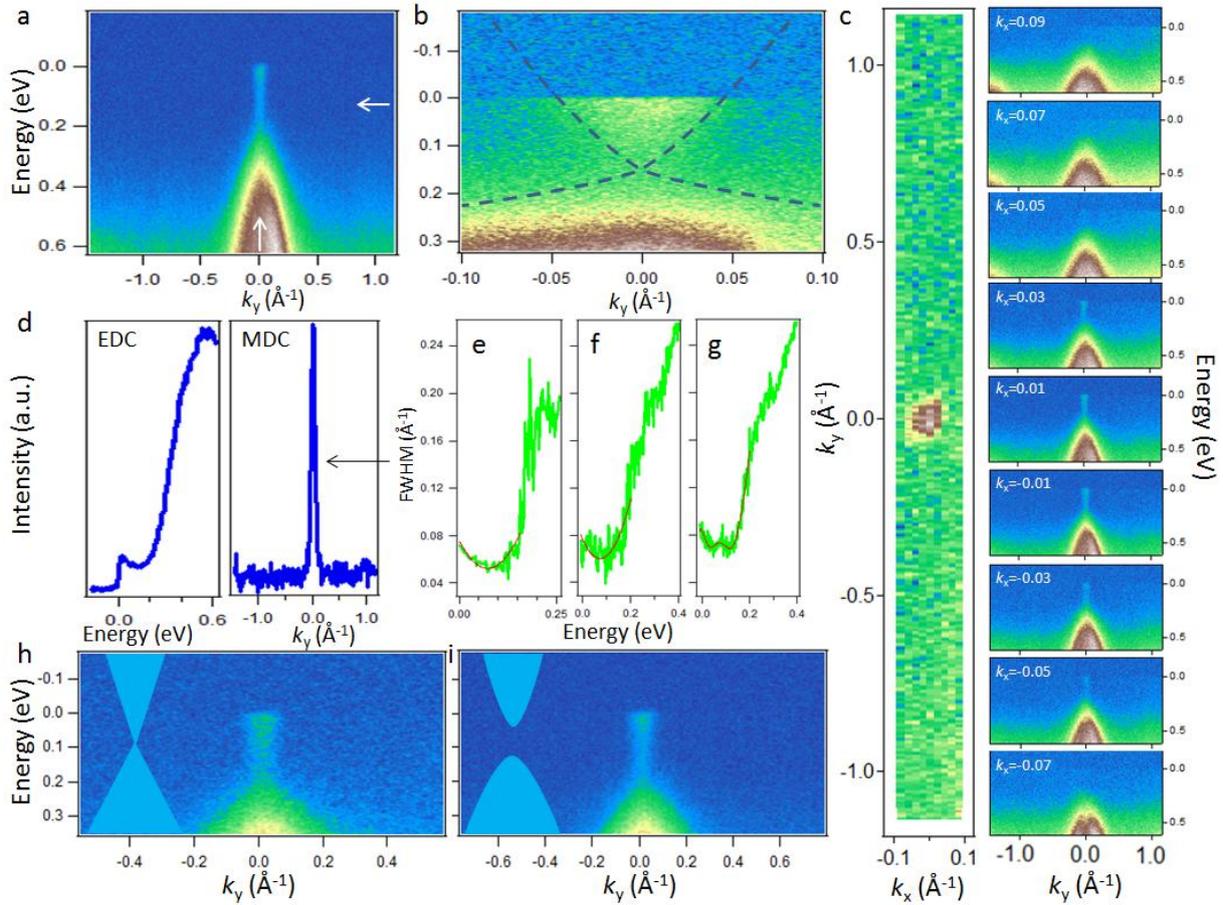

**Figure 2. In-plane and energy locations of 3D Dirac points.** (a) 90 eV data along the cut going precisely through one of the $k_x=k_y=0$ points. (b) The same dataset taken using the 18 eV photons. Dashed lines are guides to the eye going very close to one of the constant intensity profiles. (c) Fermi surface map (25 meV energy window at the Fermi level) and corresponding intensity plots for each of the $k_x$ momenta. (d) Typical energy- and momentum distribution curves (EDC and MDC) from panel (a). (e, f, g) Full-width-at-half-maximum as function of energy for the data from panels b, h and i respectively. Red solid lines are polynomial fits to the parts of the curves. (h) Intensity plot taken in normal emission geometry and (i) 0.6° away. Cartoons illustrate the underlying electronic structure integrated over $k_z$, as implied by the local minima in fits to FWHM curves from panels (e-g).

We found the suitable experimental conditions to enhance the intensity in the aforementioned region close to the Fermi level. These are the precise angular position of the sample corresponding to normal emission and a particular set of photon energies. One of such datasets taken using hv = 90 eV is presented in Fig. 2a. It clearly demonstrates the presence of strongly dispersing electronic states in the 200 meV energy window below the Fermi level. Corresponding to the white arrows in panel 2 **a,** energy- and momentum distribution curves (MDC) are shown in panel **d**. They reveal a well-defined structure that is strongly localized in momentum. In order to resolve the possible fine structure in this region, we have carried out the measurements with the highest possible (at 18 eV photon energy) resolution (~3 meV) at ~0.9 K. The result is shown in Fig. 2b. We have observed the cone-like structure, but without the single dispersions that cross each other as in the case of the surface states of topological insulators [13, 14]. This can be considered as a first hint of the presence of a 3D Dirac point, as one expects to see exactly such an intensity distribution in ARPES when $k_z$ can only be defined with significantly worse resolution than the in-plane components. A typical example of this is the bulk conduction band in $Bi_2Se_3$ [13, 14].

We took further steps to characterize the narrow cone-like electronic feature in other directions of momentum and in energy. Fig. 2c shows the Fermi surface map with its underlying momentum-energy cuts (right panels of Fig. 2c). As is seen from the map, the feature is also strongly localized along the $k_x$ direction, giving rise to a point-like Fermi surface. In order to find the energy of the possible touching of the valence and conduction bands we plot the width of the MDCs, as is shown in Fig. 2d, as a function of binding energy and present the results in Fig. 2e-g. All curves show pronounced minima, some of them even showing two. Those spectra, which are taken exactly along the cuts going through Γ (Fig. 2 b, h) have a minimum at ~75 meV from the Fermi level (Fig. 2e, f). The one taken a fraction of a degree away in orientation (Fig. 2i), shows two minima and a local maximum, again at ~75 meV (Fig. 2g). Although purely visual inspection of the data implies that the bands touch below 100 meV, we ascribe the position of the minimum of the MDC's width (at ~75 meV) to the crossing point of the 3D Dirac cone.

To identify the $k_z$ location we have scanned the photon energy in the range from 15 eV to 110 eV, covering many reciprocal lattice constants in that direction. Fig. 3 summarizes these results. Partly because of matrix elements effects and partly because of the overlap with the signal from higher orders, we managed to detect the feature only at six photon energies in this interval. We show for comparison only several datasets (panels b, c, e) from a detailed search, which do not show any evidence for the intensity connecting the top of the valence band with the Fermi level. This is in sharp contrast with the case of topological insulators, for which surface states are seen with the same clarity practically independent of photon energy. We used the free electron approximation to calculate $k_z$ values from the photon energies given in Fig.3, which is the accepted practice in the ARPES community. Using this procedure, it was not possible to find a suitable value for the inner potential, the only variable parameter in this calculation, which would attribute all "lucky" photon energies to a single $k_z$, i.e. to the Γ or Z-points. Therefore, the only reasonable assumption is that, in accordance with the calculations, the Dirac point is located somewhere in between Γ and Z. Two values of the inner potential correspond to this situation (4.3 and 8.5 eV). In spite of the apparently non-matching $k_z$ s for 27 eV and 35 eV photon energies (Fig. 3 h,i) in the case of $V_0$=8.5 eV, we included this possibility because the intensity of the feature in both cases is very weak and the corresponding $k_z$ s can thus be used to estimate the error bars. Such deviations can be caused by many factors natural to the case of employing the free electron approximation for the final state. We can thus conclude that the detected crossing of the valence and conduction bands is of bulk origin and is localized between the Γ and Z points, somewhat closer to the former. This is in close agreement with our band structure calculations as well.

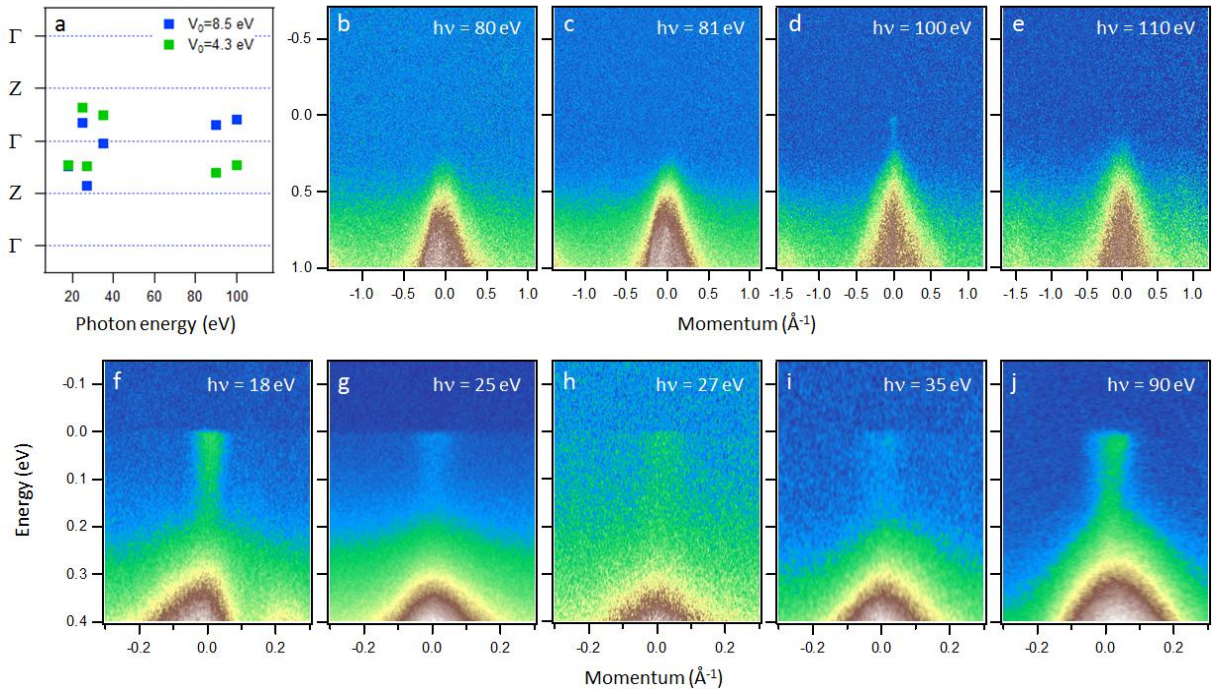

**Figure 3. Out-of-plane location of 3D Dirac points.** (a) Calculated $k_z$ values for two values of the inner potential 4.3 eV and 8.5 eV. Note, the values nearly coincide for hν=18 eV. Here ΓZ is 0.247 Å$^{-1}$, i.e. corresponding to the body centered tetragonal unit cell. (b-j) ARPES data recorded using different photon energies in fixed geometry (including normal emission). In all cases normal emission coincided with the forward direction to the analyzer.

The experimental results presented in Figs.1-3 are consistent with the presence of a single pair of 3D Dirac points in cadmium arsenide, which can thus be considered as the first real material realization of a 3D Dirac semimetal (Fig. 4a). The only reservation that should be made is the presence of the finite n-doping in the current samples, which slightly shifts the Fermi level from the Dirac points – in analogy to what is often seen with topological insulators [13]. In order to visualize the effect of doping on the electronic structure of the 3D Dirac semimetal, in Fig. 4b,c we show schematically the dispersions and Fermi surfaces for the n- and p-doped cases. An alternative way to depict a 3D Dirac point, in essence a four-dimensional object, is given in Fig. 4d. Here the energy is represented by the linear color scale which allows one to restore E(***k***) curves along an arbitrary direction in the vicinity of the 3D Dirac point. Future materials development or gating experiments are expected to bring the chemical potential down to exactly the Dirac point.

Another similarity of the current compound with known topological insulators is that $Cd_3As_2$ itself has been known since long ago [15], as were Bi-Sb alloys and $Bi_2X_3$ thermoelectrics. Earlier studies already indicated that this material is very unusual. Previous interest in $Cd_3As_2$ has been due to its very high electron mobility [6, 16] and its overall electronic properties, which have been attributed to the presence of an inverted band structure [17, 18]. Our results are fully consistent with previous experiments. Indeed, the high electron mobility observed, up to 280,000 cm$^2$/Vs [6], which is comparable to that of the best graphene [free standing graphene], can be explained considering the fact that the mobility is directly proportional to the Fermi velocity and inversely proportional to the scattering rate and Fermi momentum. As shown above, the Fermi surface of cadmium arsenide consist of two tiny ellipsoids (Fig. 2c and Fig. 4c) or almost spheres, which is in agreement with earlier Shubnikov-de Haas oscillation [19] studies. This not only implies a small Fermi momentum ($k_F$ ~ 0.04 Å$^{-1}$), but also a strongly reduced scattering rate, as the phase space for a scattering is extremely small. Knowing the electronic structure from our ARPES data, we can estimate the Fermi velocity. Because of the strong three-dimensionality, we cannot do it very precisely, as in the case of other materials where the dispersion features are very narrow in momentum, but our estimates (e.g. from the slopes of the edges of the cones clearly visible in Fig.3) give the values of 5±2 eVÅ, i.e. even higher than those for the usual forms of graphene (not free-standing). All of these factors explain the record high 3D materials' electron mobility in cadmium arsenide.

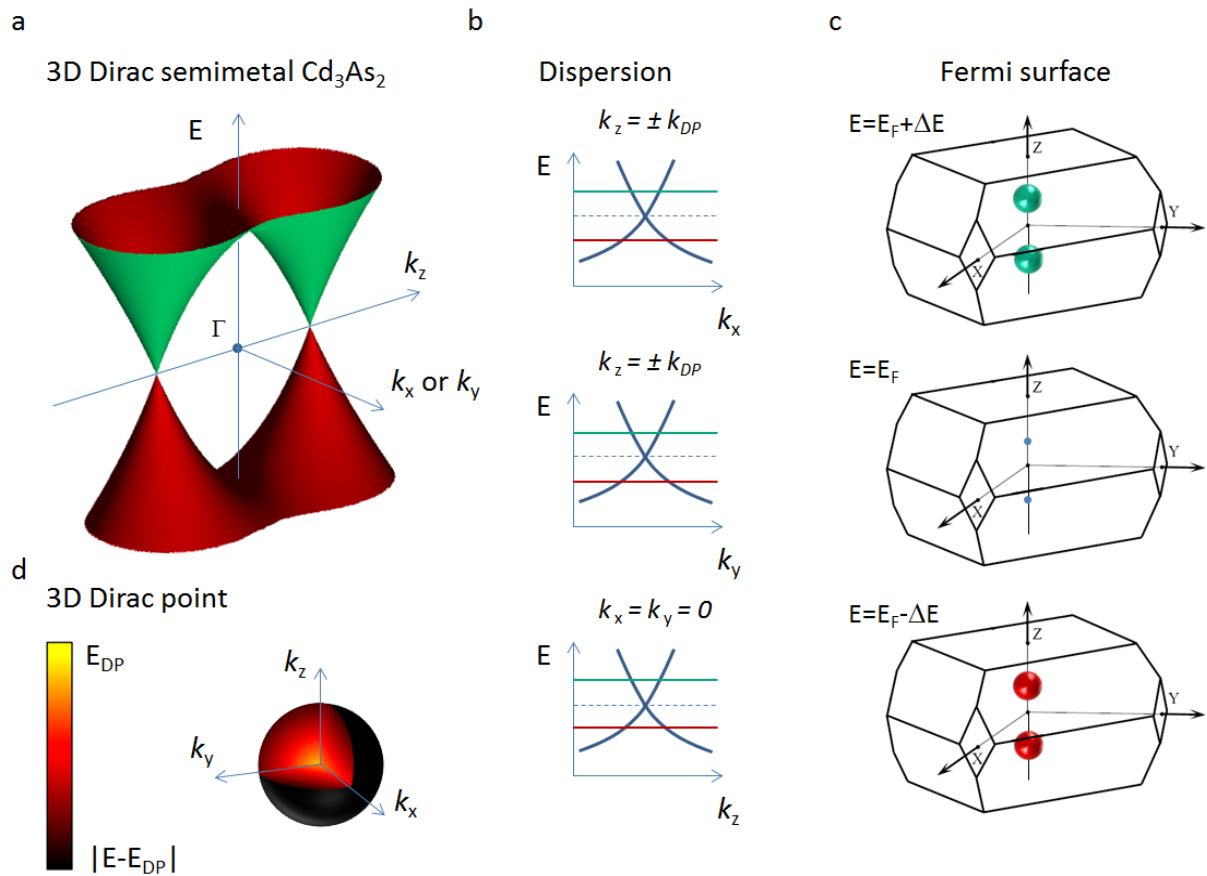

**Figure 4. Formation of a '4D-semimetal'.** (a) Schematic electronic structure of cadmium arsenide as revealed by ARPES in this study. Conduction (green from outside) and valence (red) bands are touching at two points only. (b) The crossing dispersions along any direction in the momentum space going through the Dirac point. Red and green lines correspond to n- and p-doped cases. (c) 3D Fermi surfaces corresponding to n- and p-doped (upper and lower panels) as well as undoped cases (middle panel). (d) An alternative way to depict a 3D Dirac point. Here the energy is represented by the linear color scale which allows one to restore E(k) dispersion along arbitrary direction in the vicinity of the 3D Dirac point.

It has been suggested that a Weyl semimetal can be designed on the basis of a 3D Dirac semimetal by breaking time-reversal or inversion symmetries [3-5]. Interestingly, our single crystals are noncentrosymmetric, but we cannot at this time speculate whether or not the Weyl state is displayed in $Cd_3As_2$ as we have not observed any surface states. This may be due to our limited $k_z$ resolution, as the "arcs" connecting the Weyl nodes should be seen in the $k_{xz}$ or $k_{yz}$ planes. Wang et. al. suggest that the 3D Dirac points remain degenerate even in the noncentrosymmetric structure; future studies may show this first embodiment of the 3D Dirac semimetal to display the Weyl semimetal state, either in its native state or with the breaking of time reversal symmetry.

**Methods summary**

Single crystals of $Cd_3A_2$ were grown from a Cd rich melt of composition $Cd_{0.85}As_{0.15}$. The components were sealed in an evacuated quartz tube, held at 800 °C for 1 week, and then cooled at 0.6 °C/hr to 380 °C. The tube was subsequently centrifuged at 380 °C to remove excess Cd. Phase purity was confirmed by X-ray Diffraction.

Electronic structure calculations were performed in the framework of density functional theory using the Wien2k code [20] with a full-potential linearized augmented plane-wave and local orbitals basis together with the Perdew-Burke-Ernzerhof parameterization of the generalized gradient approximation [21]. The plane wave cutoff parameter $R_{MT}K_{max}$ was set to 7 and the Brillouin zone (BZ) was sampled by 2000 k-points. Spin-orbit coupling (SOC) was included. Continuous bands were plotted by calculating the irreducible representations of the wavefunctions at each k-point, showing which of the band intersections represent symmetry-allowed crossings. For the calculations, the primitive centrosymmetric unit cell (space group $P4_2/nmc$) was used.

ARPES data were collected at Helmholtz Zentrum Berlin using the synchrotron radiation from the BESSY storage ring. Excitation photon energies in the range 15 eV – 110 eV were used under full control of polarization. The end-station "$1^3$-ARPES" is equipped with the He3 cryostat which allows one to keep the temperature of the sample at ~1K. All spectra in the present work were recorded at such temperatures. Overall energy resolution depends on the photon energy and here was set to about 3 meV at 15 eV and about 12 meV at 110 eV. Momentum resolution along ($k_y$) and perpendicular ($k_x$) to the slit the slit is ~0.2° each. All samples were cleaved at approximately 40 K.

**References**

[1] S. M. Young et al. "Dirac Semimetal in Three Dimensions", Phys. Rev. Lett. **108**, 140405 (2012)

[2] C. L. Kane and E. J. Mele, "Quantum Spin Hall Effect in Graphene", Phys. Rev. Lett. **95**, 226801 (2005)

[3] G. B. Halász and L. Balents, "Time-reversal invariant realization of the Weyl semimetal phase", Phys. Rev. B **85**, 035103 (2012)

[4] A. A. Burkov and L. Balents, "Weyl Semimetal in a Topological Insulator Multilayer", Phys. Rev. Lett. **107**, 127205 (2011)

[5] Z. J. Wang, Y. Sun, X. Q. Chen, C. Franchini, G. Xu, H. M. Weng, X. Dai, and Z. Fang, Phys. Rev. B **85**, 195320 (2012)

[6] L. Zdanowicz, J. C. Portal, W. Zdanowicz, "Shubnikov-de Haas effect in amorphous Cd3As2" in "Application of High Magnetic Fields in Semiconductor Physics", Lecture Notes in Physics **177**, 386 (1983)

[7] X. G. Wan, A. M. Turner, A. Vishwanath, S. Y. Savrasov, "Topological semimetal and Fermi-arc surface states in the electronic structure of pyrochlore iridates", Phys. Rev. B **83**, 205101 (2011)

[8] S. V. Borisenko, "One-cubed" ARPES User Facility at BESSY II, Synchrotron Radiation News **25**, 6 (2012)

[9] G. A. Steigmann and J. Goodyear, "The crystal structure of $Cd_3As_2$", Acta Crystallographica Section B: Structural Crystallography and Crystal Chemistry **24**, 1062 (1968)

[10] H. Okamoto, "The As-Cd (arsenic-cadmium) system", Journal of Phase Equilibria **13**, 147 (1992)

[11] Zhijun Wang, Hongming Weng, Quansheng Wu, Xi Dai, Zhong Fang, " Three Dimensional Dirac Semimetal and Quantum Spin Hall Effect in $Cd_3As_2$", arXiv:1305.6780

[12] S. V. Borisenko et al. „ Pseudogap and charge density waves in two dimensions", Phys. Rev. Lett. **100**, 196402 (2008)

[13] M. Z. Hasan, and C. L. Kane, "Topological insulators", Rev. Mod. Phys. **82**, 3045 (2010)

[14] A. Kordyuk et al. „Photoemission-induced gating of topological insulators", Phys. Rev. B **83**, 081303 (2011)

[15] W. Zdanowicz, and L. Zdanowicz, "Semiconducting compounds of the $A^{II}B^{V}$ group", Ann. Rev. Mater. Sci. **5**, 301 (1975)

[16] Arthur J Rosenberg and Theodore C Harman, "$Cd_3As_2$ A Noncubic Semiconductor with Unusually High Electron Mobility", Journal of Applied Physics **30**, 1621 (1959)

[17] Dowgiallo, B. Plenkiewicz, P. Plenkiewicz, "Inverted band structure of $Cd_3As_2$", Physica Status Solidi (b) **94**, K57 (1979)

[18] D. R. Lovett, "The growth and electrical properties of single crystal $Cd_3As_2$ platelets", Journal of Materials Science **7**, 388, (1972)

[19] I. Rosenman, "Effet Shubnikov de Haas dans Cd3As2: Forme de la surface de Fermi et modele non parabolique de la bande de conduction", J. Phys. Chem. Solids **30**, 1385 (1966)

[20] P. Blaha, K. Schwarz, P. Sorantin, and S. Trickey, "Full-potential, linearized augmented plane wave programs for crystalline systems", Computer Physics Communications, **59**, 399 (1990)

[21] Perdew, K. Burke, and M. Ernzerhof, "Generalized Gradient Approximation Made Simple", Phys. Rev. Lett. **77**, 3865 (1996)

**Acknowledgements**

The research at Princeton was supported by the DARPA-SPAWAR grant N6601-11-4110 and the ARO MURI program, grant W911NF-12-1-0461. RC thanks the Humboldt Foundation for support. The research in Dresden is supported by the grants BO1912/3-1, BO1912/2-2 and ZA 654/1-1.

* Present address: Physikalisches Institut, EP IV, Julius-Maximilians-Universität Würzburg, Am Hubland, D-97074 Würzburg

**Supplementary information**

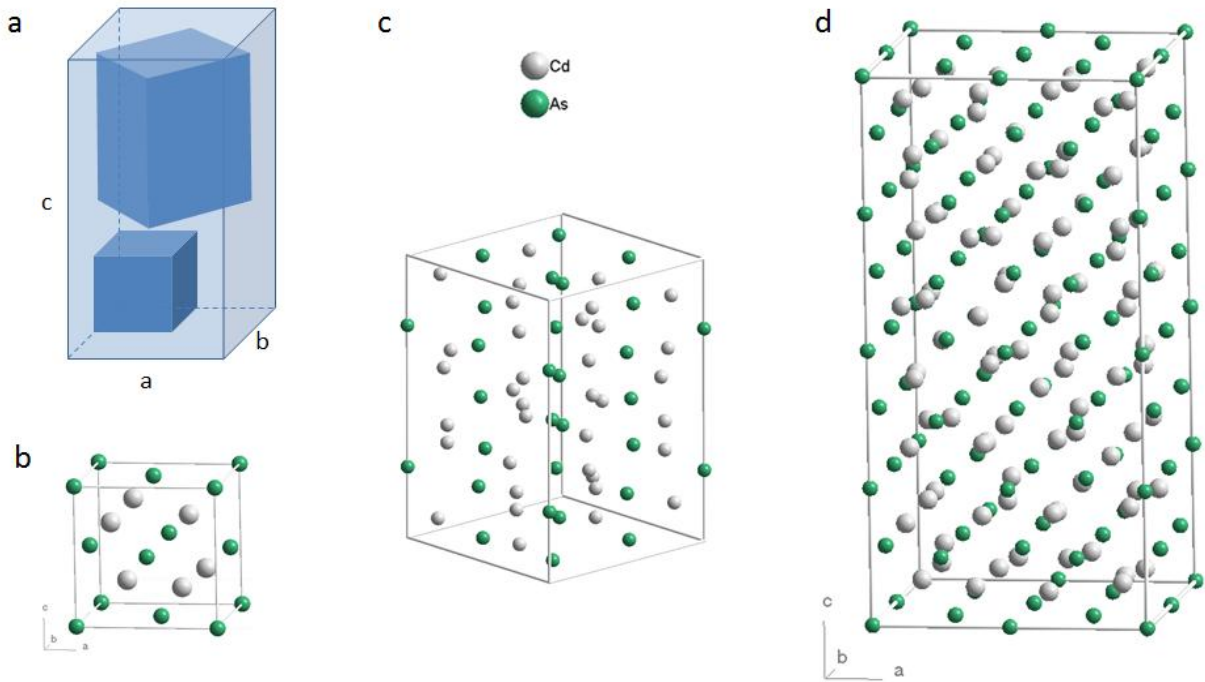

Figure S1. Crystal structure of cadmium arsenide.

In Fig. S1 The crystal structure of $Cd_3As_2$ is shown in more detail. The relation between the unit cells is shown in panel (a). The parent (nearly cubic) structure determining the most intense features of ARPES spectra is shown in panel (b). Its lattice constant is approximately 6.34 Å, which corresponds to the apparent periodicity of the order of 1Å$^{-1}$ seen in the valence band spectra (Fig. 1, Fig.S3, Fig. S5). This structure has random Cd vacancies. The primitive centrosymmetric tetragonal cell (space group $P4_2/nmc$) used in the calculations is shown in panel (c) and the low temperature non-centrosymmetric unit cell (space group $I4_1cd$) is shown in panel d.

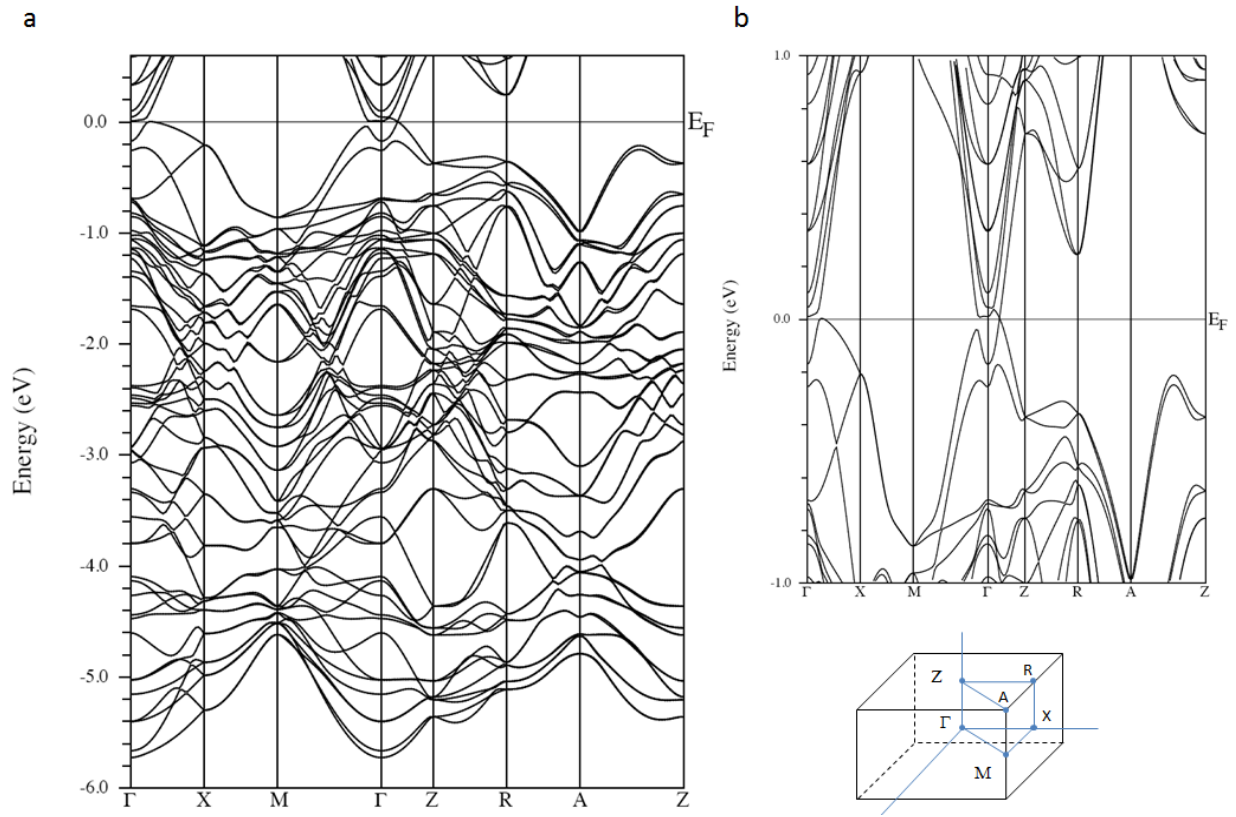

Figure S2. Band structure of the high-temperature phase of cadmium arsenide.

Complete results of the band structure calculations are presented in Fig. 2S together with the corresponding Brillouin zone. The noticeable dispersion along the ΓZ direction of majority of the states explains the relatively broad linewidth of the ARPES spectra (Fig. 1, Fig. S3, Fig. S5). Those features which disperse more moderately are seen better in the experiment, e.g. near 1eV and 4 eV binding energies.

In order to resolve the fine structure of the valence band we used light of different polarizations. Because of matrix elements effects, some of the components may be suppressed. While indeed we could see more dispersing states, e.g. the broad features running from the top of the valence band to both sides turn out to consist of three components each.

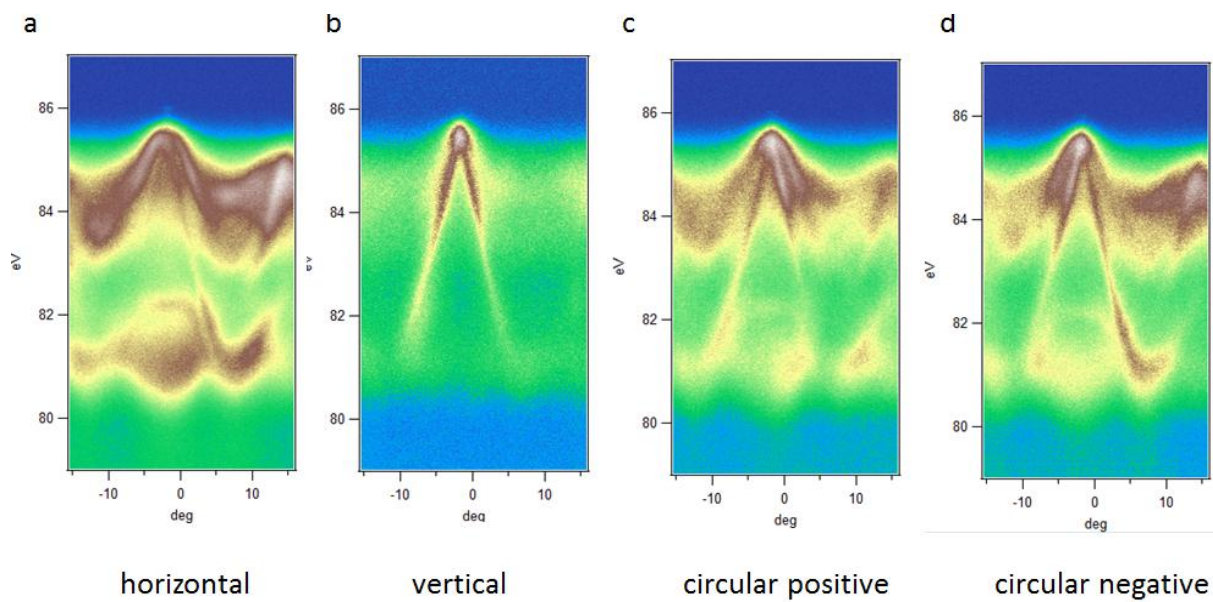

Figure S3. Valence band of cadmium arsenide measured with light of different polarizations.

We also checked the composition of the material by recording conventional XPS spectra (Fig. S4a). Characteristic peaks from core levels of cadmium and arsenic were clearly seen. In order to exclude the charging effects which may take place while measuring semiconductors, especially at such low temperatures (~1K), we have recorded some spectra using a different flux. As is seen from Fig. S4b, where the two EDCs (Energy Distribution Curves) were taken by varying intensity by the factor of 4 and no noticeable shift has been detected, the sample is a very good conductor. This confirms the semimetallic nature of $Cd_3As_2$.

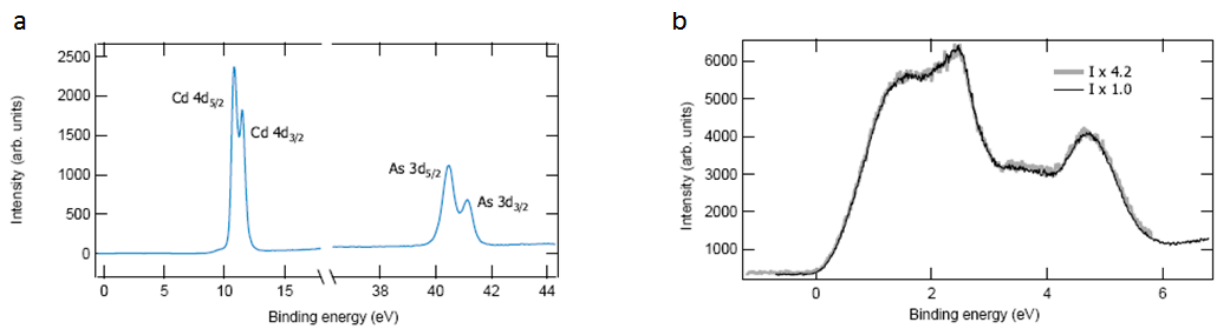

Figure S4. (a) XPS spectra of the core levels of Cd and As. (b) Absence of charging in the valence band spectra.

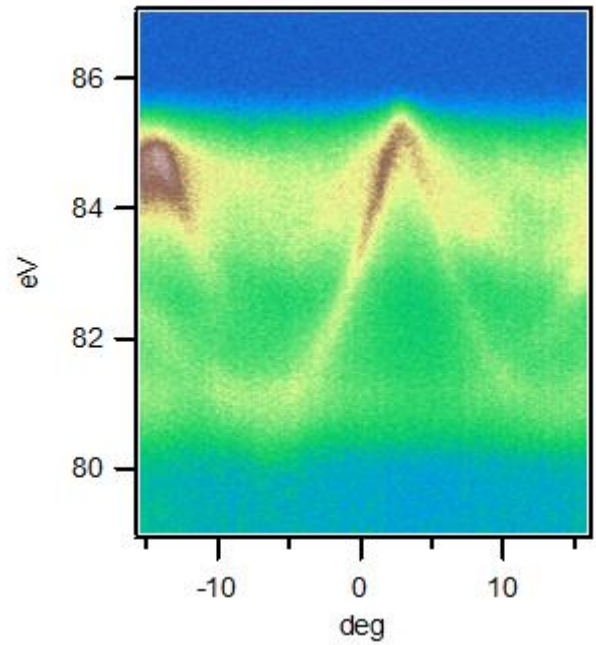

Figure S5. Spectrum from the center of the 2$^{nd}$ Brillouin zone.

Finally, in Fig. S5 we show the spectrum analogous to that from Fig. 2, but taken at an angle corresponding to the second Brillouin zone. Although the excitation energy is the same, we observed no cone-like structure at the $\Gamma$ point. This is another confirmation that we are characterizing the bulk states. Because the absolute value of the perpendicular component of the momentum depends also on the angle of emission, here we probe different $k_z$. Thus, the observed cone-like structure in normal emission is indeed localized also along $k_z$ and has nothing to do with the surface states which should have been seen here with the same clarity.